\documentstyle[aps]{revtex} 
\begin{document} 
\section*{{Reply to comment by Zaslavskii on extremal black hole action}}
\begin{abstract} 
It is shown that Zaslavskii's misunderstanding of our published proof of the
irrelevance of all extremal black hole configurations (whether with equal
charge and mass or not) rests on his refusal to see the essential difference
between the correct inequality governing extremal and non-extremal actions
and his incorrect version.
\end{abstract}

\bigskip

Recently in a comment \cite{z} on our already published reply
\cite{GMrep} Zaslavskii expressed the opinion that our relation
$I_n(m,q-\epsilon)<I_e(m,q)$ \cite{GMrep} between non-extremal and
extremal on-shell Reissner - Nordstr\"{o}m actions
is essentially the same as the
relation $I_n(m,q)<I_e(m,q)$ \cite{zcom}.  But the latter relation, which
is his misinterpretation of a statement in our Letter \cite{GM}, leads him
astray, when he relies on the {\it equality} of the charges on the two sides
at the top of p. 2 (\cite{z}:v.1). On p. 4, while commenting on the
former (correct) inequality, he forgets this and claims that the fact
{\it "that charges on both sides ... may (be) slightly different from
each other, is not crucial"}! This contradicts his use of the inequality.

Our point was that for each extremal configuration $(m,q)$, there is a
non-extremal configuration $(m,q-\epsilon)$ of lower action. This follows
trivially from the first inequality, but cannot be understood from the
second inequality (which does not exist anywhere in \cite{GMrep,GM}
and is Zaslavskii's invention) because in the case $m=q$ one would 
need a non-extremal configuration with $m=q$, but such a thing does not 
exist, whereas non-extremal configurations with $q=m-\epsilon$ of course do. 
For extremal configurations only with $m\ne q$, the second relation can be 
used. In all cases, the corresponding non-extremal configurations have 
lower action, so that extremal configurations cannot be physically relevant 
in the approximation considered.

His other comments (on, {\it e.g.}, the evaluation of path integrals under
the approximations considered in \cite{GM}) show further misconceptions
which, however, are not related to the main point involved in \cite{GM}.

\bigskip
\noindent{A. Ghosh}\\
{CERN, CH-1211 Geneva 23, Switzerland}\\
and P. Mitra\\
{Saha Institute of Nuclear Physics,
Block AF, Bidhannagar,
Calcutta 700 064, India}

\begin{references}
\bibitem{z} O. B. Zaslavskii, hep-th 9804090 (1998)
\bibitem{GMrep} A. Ghosh and P. Mitra, Phys. Rev. Letters, {\bf 80}, 3413 (1998)
\bibitem{zcom} O. B. Zaslavskii, Phys. Rev. Letters, {\bf 80}, 3412 (1998)
\bibitem{GM} A. Ghosh and P. Mitra, Phys. Rev. Letters, {\bf 78}, 1858 (1997)
\end{references}
\end{document}